\documentclass[doublecol]{epl2} 
\usepackage[english]{babel}
\usepackage{amsmath,amssymb,bbm,mathrsfs,mathtools,multirow,diagbox,xcolor,%
  graphicx,tabularx,comment,amsfonts,dsfont,lettrine,units,comment}
\usepackage{graphicx}
\usepackage{bookmark}
\usepackage{xcolor}
\usepackage{chngcntr}
\usepackage{braket}
\usepackage{nicefrac}
\usepackage{bm}
\usepackage{bbm}
\usepackage{braket}
\usepackage[normalem]{ulem} 
\usepackage{hyperref}
\hypersetup{
    colorlinks=true,
    linkcolor=blue,
    filecolor=blue,      
    urlcolor=blue,
    citecolor=blue,
    backref=page,
    pdftitle={a-review},    
    pdfkeywords={amorphous}, 
    pdfnewwindow=true,      
}

\title{Amorphous topological matter: theory and experiment}
\shorttitle{Amorphous topological matter: theory and experiment} 

\author{Paul Corbae \inst{1,2} \and Julia D. Hannukainen \inst{3} \and Quentin Marsal \inst{4} \and Daniel Mu\~noz-Segovia \inst{5} \and Adolfo G. Grushin \inst{4} \footnote{All authors contributed equally to the writing of the manuscript.} \footnote{adolfo.grushin@neel.cnrs.fr}}
\shortauthor{P. Corbae \etal}

\institute{                    
  \inst{1} Department of Materials Science, University of California, Berkeley, California 94720, USA\\
  \inst{2} Materials Science Division, Lawrence Berkeley National Laboratory, Berkeley, California 94720, USA\\
  \inst{3} Department of Physics, KTH Royal Institute of Technology, 106 91 Stockholm, Sweden\\
  \inst{4} Univ.~Grenoble Alpes, CNRS, Grenoble INP, Institut Néel, 38000 Grenoble, France\\
  \inst{5} Donostia International Physics Center, 20018 Donostia-San Sebastian, Spain
}
\pacs{nn.mm.xx}{First pacs description}
\pacs{nn.mm.xx}{Second pacs description}

\abstract{
Topological phases of matter are ubiquitous in crystals, but less is known about their existence in amorphous systems, that lack long-range order. In this perspective, we review the recent progress made on  theoretically defining amorphous topological phases  and the new phenomenology that they can open. We revisit key experiments suggesting that amorphous topological phases exist in both solid-state and synthetic amorphous systems. We finish by discussing the open questions in the field, that promises to significantly enlarge the set of materials and synthetic systems benefiting from the robustness of topological matter.}

\begin{document}

\maketitle

\section{Introduction} 

The quantum Hall state, the first topological phase ever observed, was discovered in crystalline heterostructures~\cite{vonKlitzingRMP1986}, even though its existence does not require an underlying crystalline lattice.
Indeed, a two dimensional free electron gas  under a perpendicular magnetic field displays Landau levels.
Its associated metallic topological edge states, and quantized conductance arise in a confining potential, with no assumption of an underlying crystalline lattice. 
The quantum Hall displays a continuous translational invariance, and the corresponding electron's momentum $\mathbf{p}$ enters the parabolic dispersion relation $\mathbf{p}^2/2m$, with $m$ being the electron's mass.
By promoting $m$ to the effective mass of the electron within a medium, the parabolic dispersion and its corresponding Landau levels can be thought of as arising from the long-wavelength limit of a lattice tight-binding crystalline model~\cite{Hofstadter76}. 
With this notion of translational invariance in place, the condensed matter community discovered how to dispose of magnetic fields to define topological states in crystalline systems~\cite{Haldane1988}, establishing topological phases in crystals of any dimension, irrespective of their insulating, conducting or superconducting nature~\cite{hasan2010,Qi_Zhang2011,Armitage2018}.

Topological phases do exist in the absence of long-range periodicity, as we are not forced to regularize a continuum theory using a periodic lattice.
This observation is at the heart of this perspective article.
Our goal is to summarize the recent progress made to understand how topological phases emerge on the largest class of non-crystalline systems, amorphous systems~\cite{agarwala_topological_2017,mitchell_amorphous_2018,xiao_photonic_2017,mansha_robust_2017}. 
Characterizing topology in amorphous matter, without the convenience of Bloch's theorem, has lead to the emergence of new phenomenology, unique to amorphous matter.
Topology remains largely unexplored in this class of solids, which may offer different functionalities compared to crystals.
We start by discussing the main properties of amorphous and topological matter, followed by a review of the progress made in combining these two fields.
We finish by summarizing the experimental status and offering some perspectives on the main open questions.
For a more technical review we refer the reader to Ref.~\cite{Grushin2020}.

\subsection{Basic properties of amorphous matter}
\begin{figure}
    \centering
    \includegraphics[width=\columnwidth]{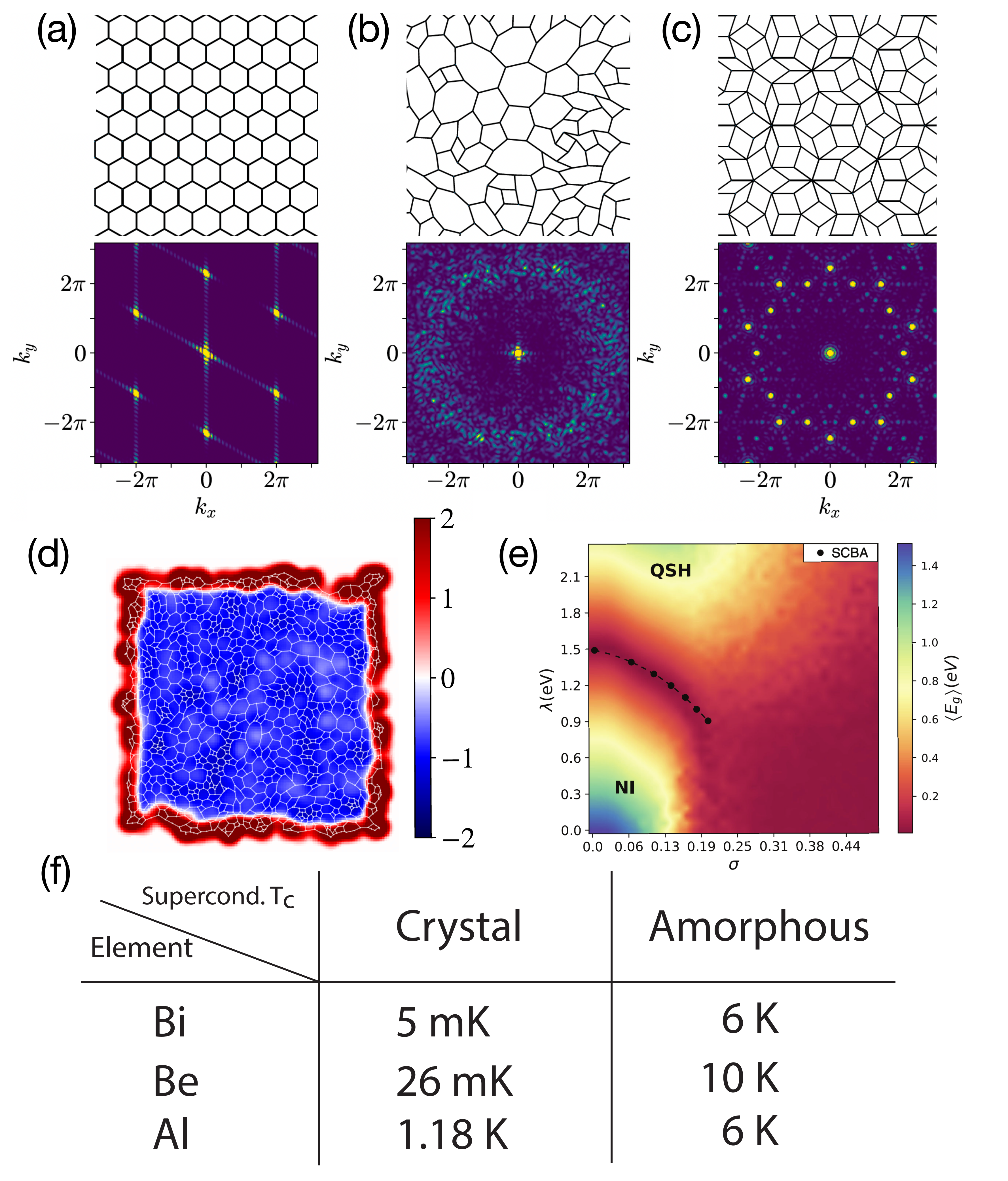}
    \caption{(a) Crystalline lattice and corresponding structure factor with sharp Bragg peaks. (b) Amorphous lattices with local order result in broad diffraction rings. (c) Quasicrystals break translational invariance but retain long range order, resulting in sharp Bragg peaks. (d) Local Chern marker of the 2D threefold-coordinated Weaire-Thorpe model~\cite{marsal_topological_2020}. The bulk average equals $-1$, indicating a nontrivial Chern insulator phase.  (e) Configuration-averaged spectral gap, $E_g$, of a 2D quantum spin Hall model on a structurally disordered trigonal lattice as a function of disorder strength $\sigma$ and spin-orbit coupling $\lambda$. The quantum spin Hall (QSH) and normal insulator (NI) phases are labelled according to the value of the spin Bott index. Adapted from Ref.~\cite{wang_structural_2022}. (f) Superconducting critical temperature of different amorphous and crystalline solids~\cite{tsuei_amorphous_1981}.} 
    \label{fig:fig1}
\end{figure}

Amorphous materials are defined by their lack of long-range order\cite{zallen_physics_1998}. However, they display short- and even medium-range order, as well defined nearest and next-to-nearest neighbour distances, respectively (Fig. \ref{fig:fig1} (a-c)). The short range order manifests itself as preferred bond lengths and angles, peaked around the values of its crystalline counterpart. Due to the short range order, amorphous materials have a well defined coordination environment with a distinct number of nearest neighbours. In solid state systems this is a result of the electronic configuration of the atoms involved in bonding. Hence, amorphous solids remain locally ordered~\cite{zallen_physics_1998, weaire_electronic_1971}.

Elucidating the atomic structure of amorphous solids is necessary to understand most of their electronic properties~\cite{zallen_physics_1998,Yangstruct2021}.
The disordered atomic positions in amorphous solids 
result in diffuse rings in the diffraction pattern and a lack of sharp Bragg peaks characteristic of crystalline materials~\cite{zallen_physics_1998}. 
The absence of discrete crystalline symmetry, in favour of local short range order and well defined diffraction rings demonstrates that amorphous systems are isotropic on average
(see Fig.~\ref{fig:fig1})~\cite{zachariansen_atomic_1932, lewis_fifty-years_2022, wright_eighty-years_2013, wright_great-crystallite_2014, porai_structure_1985}.
Fluctuations of the bond lengths account for the broadening of the rings. The radii of the diffraction rings and their weight can be used to determine the structure factor and estimate an average bond distance and coordination number of the amorphous structure~\cite{warren_summary_1941, valenkov_x-ray-investigation_1936}.
The absence of Bragg peaks in the diffraction pattern, and thus the absence of long-range order, determines which solids are amorphous.

Amorphous materials are commonplace in science and technology\cite{zallen_physics_1998}.
Their applications range from common objects such as window glass to technological devices like computer memories or solar cells~\cite{legallo_overview_2020, gaspard_structure_2016}. 
Amorphous materials are advantageous for technological applications as they can be grown under less stringent conditions than single crystals require. In solids state systems, they can be grown in a range of compositions unlike typical crystalline compounds. Transitioning between the amorphous to the crystalline state in a controlled and reversible manner, for example using current or laser pulses \cite{PhysRevB.69.104111}, is a useful and defining property of phase-change materials. These are commonly used in computer memory-storage devices~\cite{legallo_overview_2020, gaspard_structure_2016}. Additionally, amorphous materials play a major role in fundamental science e.g. as coatings in gravitational waves detectors at LIGO\cite{PhysRevD.91.062005}. 

Similarly to crystals, amorphous materials can be insulators, semiconductors, metals, and superconductors~\cite{zallen_physics_1998}. Amorphous oxides used in glassware, such as silicon oxide or lead glass, are century-old insulators.
Amorphous semiconductors, such as silicon or germanium, have also been extensively studied, due to their possible use in electronic devices~\cite{weaire_electronic_1971}. Amorphous metals are exceptionally hard and can display unique magnetic properties \cite{PhysRevLett.68.1391}. Amorphous superconductors can also be synthesised \cite{barzola-quiquia_superconductivity_2016}, as conventional superconductivity is robust to disorder, an observation known as Anderson's theorem~\cite{Anderson_theory_1959}.
Remarkably, the critical superconducting temperature has been observed to be higher in several amorphous materials compared to their crystalline counterpart (Fig.~\ref{fig:fig1}(d)).

The existence and robustness of topological phases poses the natural question of whether they can be realized in amorphous systems. Before reviewing how topological amorphous phases were first achieved~\cite{agarwala_topological_2017,mitchell_amorphous_2018,xiao_photonic_2017,mansha_robust_2017} and extended, we revisit the main properties of topological phases.

\subsection{Basic properties of topological matter}

The discovery of the quantum Hall effect and its quantized Hall conductance~\cite{klitzing1980, laughlin1981, thouless1982} introduced the field of topological matter; phases of matter characterized by their metallic boundary states and quantized responses to external fields, which are robust against impurities and local perturbations~\cite{moessner_moore_2021}.
The quantum Hall effect is an example of a strong topological insulator~\cite{hasan2010, Qi_Zhang2011}, phases of matter where the boundary states are protected by local symmetries.
These symmetries are the time reversal-,  particle hole-, and chiral symmetry, the latter being the product of the other two. These three symmetries can be combined in ten different ways, defining the Altland-Zirnbauer classification of first quantized free fermion Hamiltonians~\cite{altland97,zirnbauer96}, leading to the full classification of strong topological insulators and superconductors~\cite{kitaev09, schnyder08,Ryu_2010, Ryu2012,ludwig15}.
There are five non-trivial Altland-Zirnbauer classes, classes that can host topological phases, in every dimension, each defined by a topological invariant, either integer valued Chern or winding numbers or a $\mathbb{Z}_2$ invariant~\cite{Ryu_2010}, characterizing the phase.
Two states are defined to be in the same topological phase if they can be adiabatically perturbed into one another smoothly without closing the conduction gap and not breaking the underlying symmetries, while keeping the number of orbitals fixed during the process.

Translational invariance in crystal lattices allows use of Bloch's theorem to define crystal momentum, simplifying the characterization of the topological phases and yielding closed-form momentum-space expressions of the topological invariants.
For example, the Chern number~\cite{nakahara18} characterizing the quantum Hall phase in two dimensions (2D) is evaluated as the integral of the Berry curvature~\cite{Berry1984} over the first Brillouin zone.

Electronic topological phases extend beyond strong topological insulators and superconductors, including weak~\cite{Fu2007,Moore2007,Roy2009}-- and crystalline~\cite{Fu2011}-- topological insulators, and topological metals~\cite{Armitage2018}.
Weak topological insulators can be constructed by stacking strong topological insulators, giving rise to symmetry protected surface states perpendicular to the stacking direction.
Crystalline (point group) symmetries, including rotations and reflections, can protect topological states called higher order topological insulators which host symmetry protected states on those surfaces which are invariant under the crystalline symmetry.
Crystalline symmetries simplify how to identify topological phases, through the concept of symmetry indicators~\cite{Kruthoff17,Po:2017ci,Bradlyn2017,Song:2018cj}-- eigenvalues of point group operators whose products determine topological invariants.

Although translation invariance simplifies describing and classifying topological phases, it is not necessary for their existence. 
For example, strong topology is protected by local symmetries, irrespective of the lattice details.
Non-trivial topology only requires the existence of a mobility gap, and not a spectral gap.
However, characterizing  topological phases of matter far from the crystalline limit, notably for non-crystalline lattices, requires new tools, as the known momentum space expressions for topological invariants are no longer applicable.
We describe these tools and the models introduced to study amorphous topological matter next.

\section{Theory of amorphous topological matter} 

\subsection{Overview}
There are a variety of amorphous models displaying topological phases, ranging from strong topological states to spatial-symmetry-protected topological phases.
Amorphous strong topological states include 2D Chern insulators in class A \cite{agarwala_topological_2017,xiao_photonic_2017,mansha_robust_2017,mitchell_amorphous_2018,minarelli_engineering_2019,chern_topological_2019,marsal_topological_2020,sahlberg_topological_2020,ivaki_criticality_2020}, 2D and 3D time-reversal invariant topological insulators in class AII \cite{agarwala_topological_2017,sahlberg_topological_2020,mano_application_2019,costa_toward_2019,focassio_structural_2021,wang_structural_2022}, and 2D time-reversal breaking topological superconductors in class D \cite{poyhonen_amorphous_2018,manna_noncrystalline_2022}. 
Amorphous structures also support phases \textit{a priori} protected by crystalline symmetries, such as 2D reflection-symmetry-protected topological insulators~\cite{spring_amorphous_2021}, 2D and 3D higher-order topological insulators~\cite{agarwala_higher-order_2020,wang_structural-disorder-induced_2021,Peng2022}, 2D and 3D obstructed insulators~\cite{marsal_obstructed_2022}, and 3D topological metals~\cite{yang_topological_2019}.
While structural disorder is detrimental to some of these states, it can also induce nontrivial phases when starting from a trivial crystalline state~\cite{wang_structural_2022,yang_topological_2019,wang_structural-disorder-induced_2021}, and it can give rise to new phenomenology intrinsically associated with amorphous topological matter and phase transitions~\cite{yang_topological_2019,sahlberg_topological_2020,ivaki_criticality_2020,spring_amorphous_2021,marsal_obstructed_2022}. 

A common starting point is a crystalline tight binding Hamiltonian known to host a topologically nontrivial phase.
The hopping terms are generalized to account for arbitrary angles and distances between sites.
For example the angular dependence can be modelled using the Slater-Koster parametrization \cite{slater_simplified_1954}, and the readial dependence can be accounted for by an exponential~\cite{agarwala_topological_2017,mukati_topological_2020,agarwala_higher-order_2020,wang_structural-disorder-induced_2021,yang_topological_2019,manna_noncrystalline_2022,spring_amorphous_2021,sahlberg_topological_2020,ivaki_criticality_2020} or polynomial~\cite{wang_structural_2022} decay with the radial distance.
There are several ways to introduce structural disorder, including lattices with uncorrelated random sites~\cite{agarwala_topological_2017,mukati_topological_2020,agarwala_higher-order_2020,wang_structural-disorder-induced_2021,poyhonen_amorphous_2018,sahlberg_topological_2020,ivaki_criticality_2020,yang_topological_2019,manna_noncrystalline_2022,spring_amorphous_2021}, more realistic models which preserve the local coordination number~\cite{mitchell_amorphous_2018,marsal_topological_2020,marsal_obstructed_2022,minarelli_engineering_2019}, and 
lattices with controllable deviations from the crystalline limit~\cite{wang_structural_2022,wang_structural-disorder-induced_2021,xiao_photonic_2017,mansha_robust_2017}.

\subsection{Characterizing topology without translational symmetry}
Among the different methods to characterize topological phases far from translationally invariant limits topological markers are a wide-spread tool.
Topological marker is a unifying term that includes the local markers~\cite{Kitaev20062, bianco11, Hugues2019, Irsigler2019, Hannukainen2022, chen_universal_2022,Ornellas2022,guzman_geometry_2022}, the spectral localizers~\cite{loring_spectral_2020,schulz-baldes_spectral_2021,cerjan_local_2022,cerjan_operator-based_2022}, the nonlocal (spin) Bott indices~\cite{Prodan2010, Prodan2011, Loring2010, LORING2015,Huang2018b,Huang2018, Loring2019, Jezequel2022, yang_topological_2019,focassio_amorphous_2021}, and similar generalizations of the winding of the quadrupole and octupole moment~\cite{agarwala_higher-order_2020,wang_structural-disorder-induced_2021}. 
Markers characterizing the two-dimensional quantum Hall phase are especially well explored, including the local Chern marker~\cite{Kitaev20062,bianco11} and the nonlocal Bott index~\cite{Loring2010}.
The local Chern marker~\cite{Kitaev20062,bianco11} is the Fourier transform of the Chern character.
For a crystalline lattice it quantizes to the Chern number at each lattice point. 
For non-crystalline lattices quantization requires averaging over a large enough region, where the size of the region is model dependent~\cite{bianco11,Hannukainen2022}, (see Fig.~\ref{fig:fig1}(d)).
The chiral and Chern-Simons markers~\cite{Hannukainen2022} are local markers analogous to the Chern marker in odd dimensions.
The chiral marker characterizes the $\mathbb{Z}$ invariant topological phases with chiral symmetry, whilst the Chern-Simons marker characterizes $\mathbb{Z}_2$ invariant phases with either time reversal or particle-hole symmetry, depending on the dimension.
Besides the topological markers, there also exist single $\boldsymbol{k}$-point formulas to determine the Chern and spin Chern numbers \cite{ceresoli_orbital_2007,favata_single-point_2023}.

Topological states often display a characteristic transport or electromagnetic response, such as quantized longitudinal conductance, the Hall conductivity, and the Witten effect~\cite{witten_dyons_1979,wilczek_two_1987,rosenberg_witten_2010,mukati_topological_2020}, which can also be used to characterize the topological phase.
The local markers in Refs.~\cite{marrazzo_locality_2017,Ornellas2022} is for example based on the local Hall conductivity measured in the bulk of the system.
Alternatively, the scattering matrix can determine topological indices without relying on the Hamiltonian eigenstates~\cite{fulga_scattering_2012}.

Topological phases can be detected by the presence of anomalous boundary states in the local density of states calculated with open boundary conditions~\cite{hasan2010,Qi_Zhang2011}. 
Neural networks can also detect non-trivial topology, by efficiently learning features associated with topology from for example the wavefunctions~\cite{mano_application_2019}, and the flow of the entanglement spectrum~\cite{uria-alvarez_deep_2022}.
Other approaches include the effective Hamiltonian~\cite{varjas_topological_2019,marsal_topological_2020}, symmetry indicators~\cite{marsal_topological_2020}, and the structural spillage~\cite{spillage_2022}, which take advantage of the gap closing and band inversion in a topological phase transition. 
The effective Hamiltonian $H_{\mathrm{eff}}$ is defined as the inverse of the Green's function of the system projected into plane waves~\cite{varjas_topological_2019,marsal_topological_2020}.  
If the spectral gap of the total Hamiltonian closes, so does the spectral gap of $H_{\mathrm{eff}}$, allowing the detection of topological phase transitions.
Therefore, one can construct topological invariants defined in terms of $H_{\mathrm{eff}}$, which only change when the full Hamiltonian undergoes a phase transition. 
Some amorphous models display average local symmetries which are used to construct symmetry indicators based on the symmetry properties of the filled states~\cite{marsal_topological_2020}.
The structural spillage is a topological indicator that measures the amount of band inversion between an amorphous system and a crystal~\cite{spillage_2022}, where the knowledge of the topological state of the crystal is used to determine the topology of the amorphous system.

\subsection{Amorphous models with strong topology} 
The Chern insulator was the first amorphous topological phase to be characterized~\cite{agarwala_topological_2017, xiao_photonic_2017,mansha_robust_2017, minarelli_engineering_2019,chern_topological_2019,marsal_topological_2020}. 
Ref.~\cite{agarwala_topological_2017} introduced a random lattice implementation of a model that displays a Chern insulator phase on a square lattice.
The random lattice exhibits a gapped topological phase characterized by a nontrivial Bott index, edge states, and a quantized longitudinal conductance, which are all hallmarks of a Chern insulator, where the nontrivial phase is separated from trivial atomic insulators by bulk gap closings.
There exists a similar random lattice implementation of a quantum Hall state, but in the presence of a magnetic field~\cite{bourne_non-commutative_2018}.
The three- and fourfold-coordinated Weaire-Thorpe amorphous lattices~\cite{weaire_electronic_1971} with complex intrasite hoppings~\cite{marsal_topological_2020}, provide a more realistic model for covalently-bonded amorphous solids.
The local symmetries of these models makes it possible to compute symmetry indicators analogous to the ones defined for crystals \cite{Wieder22}.
These symmetry indicators predict a Chern insulator phase, which is confirmed by the presence of edge states, the nontrivial local Chern marker, (see Fig.~\ref{fig:fig1}(d)), and the effective Hamiltonian \cite{marsal_topological_2020}. 
Amorphous Chern insulators are also present in artificial systems, such as mechanical metamaterials~\cite{mitchell_amorphous_2018}, gyromagnetic photonic lattices~\cite{xiao_photonic_2017,mansha_robust_2017,Wulles2022,Skipetrov2022}, and magnetic impurities on the surface of topological insulators~\cite{minarelli_engineering_2019}.
The Chern insulator phase also survives in an atomic liquid, defined via tight-binding molecular dynamics,
which not only lacks long-range order, but has thermally moving atoms~\cite{chern_topological_2019}. Under an external magnetic field, Ref.~\cite{sahlberg_quantum_2023} realized the quantum Hall effect arising from Landau levels in a system whose microscopic lattice is amorphous.

Amorphous quantum spin Hall insulators~\cite{agarwala_topological_2017,costa_toward_2019,focassio_structural_2021,wang_structural_2022,uria-alvarez_deep_2022,Junyan2022} are characterized by a nonzero spin Bott index and edge states carrying a quantized $2e^2/h$ conductance. 
Ref.~\cite{agarwala_topological_2017} realized a quantum spin Hall phase by placing the Bernevig-Hughes-Zhang model~\cite{bernevig_quantum_2006} on a random lattice.
Ref.~\cite{uria-alvarez_deep_2022} studied a similar model and calculated its topological phase diagram using a neural network algorithm that learns the flow of the entanglement spectrum.
Refs.~\cite{costa_toward_2019,focassio_structural_2021} performed a realistic modelling of amorphous monolayer Bismuth using density functional theory, showing that the topology of the crystal survives in the amorphous structure.
Based on both tight-binding and density functional theory calculations, Ref.~\cite{spillage_2022} showed that the amorphous Bismuth bilayer remains topological, as indicated by the structural spillage and the conductance. 
Ref.~\cite{wang_structural_2022} demonstrated a structural-disorder-induced quantum spin Hall phase, constructing a phase diagram as a function of spin-orbit coupling and disorder strength, by modelling the disorder by Gaussian deviations from an initial triangular lattice. 
For a range of parameters where the initial crystal is a trivial insulator, the disorder decreases the bulk gap and favours the topological phase (see Fig.~\ref{fig:fig1}(e)), as happens in the onsite-disorder-driven topological Anderson insulators \cite{li_topological_2009,groth_theory_2009}.

Amorphous structures also display 3D time-reversal-invariant topological insulators ~\cite{agarwala_topological_2017,mano_application_2019,mukati_topological_2020}. 
Ref.~\cite{agarwala_topological_2017} described a 3D random lattice model with exponentially-decaying hoppings that, for appropriate onsite energy $M$ and range of the hopping $r_0$, displays surface states. 
Ref.~\cite{mukati_topological_2020} further characterized the $r_0 - M$ phase diagram of the same model, and found that the phase with surface states features the Witten effect---due to the axion electromagnetic term in the action, a magnetic monopole binds a half-odd integer electric charge, forming a dyon~\cite{witten_dyons_1979,wilczek_two_1987,rosenberg_witten_2010}. 
Moreover, Ref.~\cite{mano_application_2019} studied a discrete random lattice typical of quantum percolation theory: a 3D cubic lattice with nearest neighbor hoppings whose sites are occupied with a given probability $p$, which controls the number and size of vacancies. 
The analysis of the zero-energy wavefunctions with a convolutional neural network show that the topological insulator survives until $p \sim 0.5$.

Finally, Refs.~ \cite{poyhonen_amorphous_2018,manna_noncrystalline_2022} have reported gapped time-reversal-breaking 2D amorphous topological superconductors in class D. 
Ref.~\cite{poyhonen_amorphous_2018} studied a Shiba glass, an ensemble of randomly distributed magnetic moments on a gapped superconducting surface with Rashba spin-orbit coupling. 
Analogously to the topological superconducting phases induced by the subgap Yu-Shiba-Rusinov states in periodic arrays of magnetic atoms \cite{choy_majorana_2011,nadj-perge_proposal_2013,rontynen_topological_2015}, the random Shiba glass effectively realizes a 2D $p_x+ip_y$ chiral topological superconductor with nontrivial Chern number and quantized thermal conductance \cite{poyhonen_amorphous_2018}. 
In contrast to the long-range pairing in this system, Ref.~\cite{manna_noncrystalline_2022} has realised this topological superconductor in 2D Dirac models with local pairing when implemented not only in random lattices, but also in quasicrystalline and fractal lattices.

\subsection{Spatial-symmetry-protected topological amorphous models}

Amorphous systems support and induce topological phases beyond strong topological states, including systems protected by spatial symmetries.~\cite{spring_amorphous_2021,agarwala_higher-order_2020,wang_structural-disorder-induced_2021,marsal_obstructed_2022,yang_topological_2019}.
The appearance of these phases is related to the concept of statistical topological insulators \cite{fu_topology_2012,fulga_statistical_2014,diez_extended_2015,song_quantization_2015}, which are spectral insulators protected by an average symmetry. 
They display gapless boundary states pinned to the critical point of a topological phase transition, and protected from localization by the average symmetry.

Based on this idea, Ref.~\cite{spring_amorphous_2021} has 
classified all 2D amorphous statistical topological insulators protected by the average continuous rotation and reflection symmetries present in amorphous matter. 
Unlike in crystals, where reflection-symmetry-protected topological insulators display edge states only on the boundaries respecting the symmetry, their amorphous counterparts show delocalized boundary states at all edge terminations. 
Furthermore, they are characterized by a bulk $\mathbb{Z}_2$ topological invariant that can be defined from the effective Hamiltonian.

Higher order topological insulators are another example of topological insulators protected by combinations of crystalline and discrete onsite symmetries, whose amorphous counterparts have also been reported \cite{agarwala_higher-order_2020,wang_structural-disorder-induced_2021,Peng2022}. 
First, Ref.~\cite{agarwala_higher-order_2020} showed that a 2D (3D) chiral-symmetry-protected higher order topological insulator with 0D corner states is robust against bulk structural disorder as long as the boundaries remain crystalline, as indicated by the quantized quadrupolar (octupolar) moment.
Ref.~\cite{peng_density-driven_2022} extended these chiral-symmetry-protected higher order topological insulators to fully amorphous lattices, and demonstrated that they can be induced from a trivial state by varying the density of sites.
Then, Ref.~\cite{wang_structural-disorder-induced_2021} realized a structural-disorder-induced 3D higher order topological insulator with chiral hinge modes, characterized by a quantized longitudinal conductance $2e^2/h$ and a quantized  winding number of the quadrupole moment with respect to an applied magnetic flux.

Obstructed atomic insulators are a class of insulators that are topologically trivial, in the sense of being described by exponentially localized and symmetric wavefunctions, but are not adiabatically connected to the trivial atomic limit \cite{song_densuremath-2-dimensional_2017,rhim_bulk-boundary_2017,bradlyn_topological_2017,benalcazar_quantization_2019,schindler_fractional_2019,cano_topology_2022,xu_filling-enforced_2021}. 
The simplest example is the half-filled inversion-symmetric Su-Schrieffer-Heeger chain \cite{su_solitons_1979}. These examples suggest that an average Peierls-like dimerization can give rise to obstructed phases, which has been exploited by Ref.~\cite{marsal_obstructed_2022} to realize amorphous obstructed insulators.
Ref.~\cite{marsal_obstructed_2022} suggested that phase-change materials, whose amorphous form can exhibit an average dimerization characterized by a double-peak structure in the three-particle correlation function \cite{gallo_overview_2020}, can controllably realize an obstructed amorphous phase. 
The main experimental signature of amorphous obstructed insulators, which 
differentiates them from their crystalline counterparts, is the appearance of a flatband of fractional charges at all terminations, not only at the corners.

Finally, there are amorphous generalizations of Weyl semimetals, dubbed a topological amorphous metal~\cite{yang_topological_2019}.
In crystals their topological charge can be measured by the Chern number of a surface enclosing the node in momentum space~\cite{Armitage2018}. 
Ref.~\cite{yang_topological_2019} defined the amorphous counterpart based on a known time-reversal-breaking two-band Weyl semimetal model defined on a random lattice. 
The topological amorphous metal is signaled by the nonzero Bott index and Hall conductivity in the planes perpendicular to the Weyl node separation in the crystal, and by the boundary states at these planes. 
Furthermore, in contrast to its crystalline version, the topological amorphous metal displays diffusive metallic behaviour.

\subsection{Amorphous topological phase transitions} 

The critical theory of topological quantum phase transitions has been extensively studied for disordered systems, especially for the quantum Hall plateau transitions \cite{huckestein_scaling_1995,evers_anderson_2008}. 
The standard theory postulates that the transition is of the Anderson localization type, characterized by a two-parameter scaling and a diverging localization length with universal critical exponent $\nu$. 
However, recent theoretical and numerical works point to a marginal scaling with non-universal effective critical exponents, which could explain the model-dependence of $\nu$ \cite{zirnbauer_integer_2019,dresselhaus_numerical_2021}.

Motivated by the different nature of the disorder and of the driving parameter of the transition, Refs.~\cite{sahlberg_topological_2020,ivaki_criticality_2020} numerically analyzed amorphous systems.
In particular, they considered both continuum 2D random geometries as well as discrete (square and triangular) 2D lattices with randomly occupied sites, as studied in percolation theory. 
In their models, a Chern insulator in class D appears above a critical density, dependent on the parameters of the Hamiltonian. 
They examined the critical scaling of both the Chern number and the conductance, as well as the conductance distribution curves. 
While their analysis is compatible with the standard two-parameter scaling form, the localization length critical exponent $\nu$ is highly non-universal. 
The exponents interpolate between a geometric classical percolation transition~\cite{stauffer_introduction_2017} and a standard Anderson localization transition~\cite{evers_anderson_2008}.
While these differences with standard theory of disordered systems remain to be fully understood, it is possible that changing the density of sites introduces a variable length scale that modifies the range of the geometric correlations in the system, which are believed to affect the critical exponents of the transition \cite{cain_integer_2001,sandler_correlated_2004,gruzberg_geometrically_2017,chen_effects_2019,klumper_random_2019}.

\subsection{Strongly interacting amorphous topological models}
All the above phases concern amorphous but non-interacting systems. The first step towards topological amorphous many-body systems was taken by Prodan~\cite{Prodan2019}, who defined toric code models, which display topological order with anyonic excitations, in random triangulations. The groundstate degeneracy and anyonic excitation survive amorphization, even if some commutation relations of the Hamiltonian terms are modified. 

Electron-electron interactions could also lead to many-body amorphous topological phases. However, identifying these phases is challenging due to the lack of local topological markers for interacting systems.
Ref.~\cite{Kim2022} circumvented this issue by solving an amorphous Chern insulator model~\cite{agarwala_topological_2017} with strong Hubbard interactions using a parton construction.
Fractionalizing the electron into a neutral fermion $f$ and a charged boson $b$ lead to a mean-field phase diagram with a phase displaying protected electrically neutral chiral edge modes of $f$, dubbed the fractionalized amorphous Chern insulator. 
Recently, it was shown that the Kitaev spin-liquid is exactly solvable in a three-fold coordinated amorphous lattice~\cite{Cassella2022}, which survives even if the lattice is not fully amorphous~\cite{Grushin2022}. Contrary to the Kitaev Honeycomb model~\cite{Kitaev:2005ik}, which realizes a gapless spin-liquid, the amorphous model groundstate is a gapped chiral spin-liquid, featuring chiral majorana edge modes. This opens the tantalising possibility to engineering disorder, for example by focused-ion beam irradiation, to induce a chiral quantum spin-liquid without magnetic fields~\cite{Grushin2022}. 

\section{Experimental status of amorphous topological matter}

\begin{figure}
\centering
  \includegraphics[width=0.89\columnwidth]{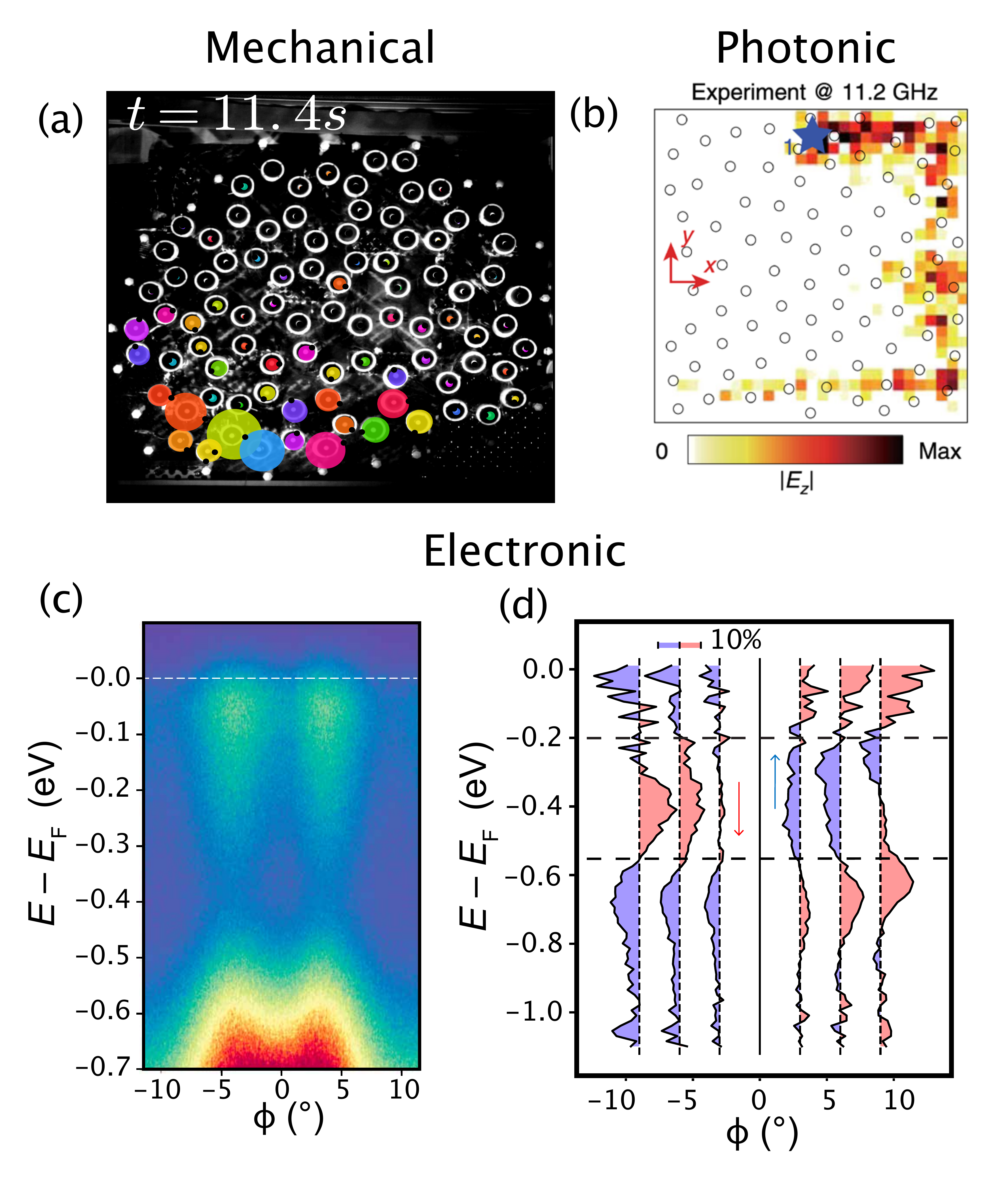}
  \caption{(a) A mechanical amorphous Chern insulator. Coupled gyroscopes excited at the edge result in a chiral edge mode. Adapted from \cite{Mitchell2018arxiv}. (b) A photonic amorphous Chern insulator with excited chiral edge modes. Adapted from \cite{Zhou2020}. (c) ARPES spectrum of amorphous Bi$_2$Se$_3$ showing well-defined dispersive features crossing the bulk gap. These midgap states ($-0.2>E-E_F>-0.6$) are two-dimensional and spin-polarized (d). The switch in polarization at $E-E_F<-0.6$ and $E-E_F>-0.2$ is consistent with bulk states.
  Adapted from \cite{corbae_evidence_2020}.}
  \label{fig:exp}
\end{figure}
Amorphous topological matter has been experimentally studied in both synthetic and solid-state systems. 
The first experimental observation was reported in a mechanical system of coupled gyroscopes \cite{mitchell_amorphous_2018}.  
Later on, the observation of spin-momentum locked surface states was reported in an amorphous electronic system \cite{corbae_evidence_2020}, as well as topological edge states in an amorphous photonic lattices \cite{Liu2020,Zhou2020,Jia2022}.

Despite the few experimental observations of topological states in amorphous matter, amorphous phases of topological matter have been frequently studied. 
In solid state systems, amorphous phases of topological materials have been studied both before and after the discovery of the quantum spin Hall effect~\cite{PhysRevLett.95.226801}. 
However experimental studies of amorphous materials did not address the survival of topological properties. 
For example, phase-change materials have been studied extensively, with GeSb$_2$Te$_4$ being one of the most widely studied representative \cite{PhysRevB.69.104111}. Interestingly, GeSb$_2$Te$_4$is also a topological insulator in its crystalline phase \cite{Nurmamat2020}. Amorphous Bi$_2$Se$_3$ has also been studied long before it was predicted to be a 3D topological insulator in its crystalline form \cite{TAKAHASHI1984261}. Amorphous and structurally disordered counterparts of crystalline topological materials have provided materials systems that show large spin-orbit torque efficiencies~\cite{doi:10.1063/1.5124688,DC2018}, but the existence and role of topological surface states have not been explored. Skyrmions, which have topologically distinct spin textures, have also been observed in amorphous systems \cite{https://doi.org/10.1002/adma.202004830}.

Using fixed-coordination amorphous structures of coupled gyroscopes, generated from different point sets such as hyperuniform or jammed, Mitchell, {\it et al.} \cite{mitchell_amorphous_2018,PhysRevE.104.025007} showed the existence of a mechanical amorphous Chern insulator with chiral, propogating edge modes, Fig. \ref{fig:exp}(a). The authors used d.c. motors that interacted via a magnetic interaction, finding that the local connectivity, which is predictive of the global density of states, is crucial for the existence of topological states in amorphous systems. Similar findings were reported in photonic systems\cite{Liu2020,Zhou2020}. By placing an amorphous arrangement of gyromagnetic rods into a waveguide and biasing them with a magnetic field, the authors observe photonic topological edge states Fig. \ref{fig:exp}(b). Interestingly, topological states exist while the system has short range order, and disappear at the glass-to-liquid transition\cite{Zhou2020}. Moreover, lattice disorder~\cite{Jia2022} enhances light confinement increasing the generation rate of correlated photon pairs by an order of magnitude compared to periodic topological platforms.

Regarding electronic materials, physical vapor deposition (PVD) is a particularly useful growth technique for amorphous materials and has been found to make amorphous materials which are not available by liquid quenching.
PVD has several advantages since it allows to control a variety of different properties, such as the substrate temperature, growth rate (which affects the time absorbed atoms have to diffuse to ideal positions), irradiation, and chemical dopants to frustrate crystallization. Modifying the substrate temperature enables the growth of amorphous films with different local ordering and produces what is called an "ideal glass" \cite{PhysRevLett.113.025503,doi:10.1073/pnas.1421042112}. 

Growth conditions are critical for achieving high quality amorphous films, especially amorphous topological materials. Growing the amorphous phase of a known topological crystal does not always preserve topological properties. Several groups have grown amorphous counterparts of known crystalline topological insulators finding no evidence for topological surface states, but rather a highly insulating, localized state \cite{Korzhovska2020,doi:10.1063/1.4908007}. First-principles calculations indicate that the local environment plays an important role in the electronic structure in three dimensional solids \cite{PhysRevB.104.214206}. If the disorder associated with the new atomic positions (new atomic environment) closes the mobility gap, the system can be trivial. 
These subtleties might explain why some amorphous versions of known crystalline topological insulators do not display evidence for a topological bulk. Controlling the growth conditions may enable tuning of the local amorphous structure. For example, by controlling the growth rate, atoms can be deposited with enough time to diffuse to their preferred neighbour (based on the chemistry of elements involved) before the next monolayer is deposited. This leads to a well defined local environment and subsequent electronic structure.

Focusing on electronic systems, the first demonstration of topological properties in an amorphous solid-state system was inspired by a known crystalline topological insulator. Bi$_2$Se$_3$ is a textbook topological insulator with quintuple layers separated by a van der Waals gap. Using PVD, Corbae, \textit{et al.}~\cite{corbae_evidence_2020} grew amorphous Bi$_2$Se$_3$ thin films with short and medium range order (next-to-nearest neighbours) as well as no van der Waals gap. In transport measurements, the films showed an increased bulk resistance that was largely temperature independent, and the weak-antilocalization effect resulting from quantum interference in two dimensions in the presence of spin orbit coupling. Using, ARPES/SARPES the authors showed that two dimensional surface states cross the bulk electronic gap and are spin polarized. The spin polarization switches multiple times as a function of binding energy matching the spin resolved spectral function from an amorphous topological model. These results contrast data taken on nanocrystalline samples which show a lack of disperson in ARPES and an insulating resistivity, consistent with earlier works \cite{PhysRevB.94.165104}. Amorphous Bi$_2$Se$_3$ in this study possesses a local environment similar to the crystal, as seen in Raman measurements, suggesting that by preserving a similar local environment to that of the crystal the topological bulk mobility gap is not closed preserving the topological nature. In contrast, the atomic environemnt at grain boundaries in nanocrystalline systems is quite disordered, providing a possible explanation for the absence of topological features. 

Looking forward, developing a workflow from growth to measurement will help the experimental observation of amorphous topological states in the solid state. In-situ measurement capabilities greatly enable spectroscopic meausurements, as they do not require thin film capping. Scanning tunneling microscopy (STM) can directly measure the electronic and real space structure. 
Combined with ARPES and transport, STM would be invaluable to discover amorphous topological materials and shed light on the nature of grain boundaries in polycrystalline~\cite{osmic_thermopower_2022} and nanocrystalline systems. 
Nano-ARPES is also promising as beam sizes scale down to the order of hundreds of nanometers.

\section{Perspective and open questions}

The growing field of topological phases in amorphous matter is an opportunity to establish a deeper understanding of topological phases and the systems that host them. In particular, the quest to define real-space topological markers and invariants to characterize topological phases is an ongoing quest. 
Defining topological indicators that signal non-crystalline topological metals remains an open question. Specifically, generalizations of Weyl semimetals to amorphous systems that respect time-reversal symmetry cannot be described by the Bott index or the Chern marker, and thus require the development of new tools.

An important open question is the lack of experimental evidence for solids that are both amorphous and topological, relating to the theoretical challenge of how to efficiently find them. The field would benefit from a \textit{textbook} amorphous topological material, where topology is unambiguously confirmed by combining different experiments. However, we lack a criterion with which to establish a hierarchy of amorphous materials where to find topological phases. 
Currently, we draw from criteria applicable to crystals, such as large spin-orbit coupling. 
However, this methodology precludes reaching the major milestone of finding materials that are only topological when grown amorphous, and that are otherwise trivial crystals.
Material candidates include amorphous Sb$_2$Se$_3$, hosting rich electronic properties as pressure changes the local environment~\cite{PhysRevLett.127.127002}, and BiTeI, predicted to host a structural topological phase transition\cite{Corbae2021}.
A promising possibility is to integrate methods such as the structural spillage and symmetry indicators, with realistic molecular dynamics predictions based on first principle calculations. 
Developing these may establish a pipeline to manufacture candidate material databases which can guide experiments. 

A related open problem is the prediction of an amorphous topological superconductor beyond toy models. Such an achievement could widen the search for platforms useful for topological quantum computing. While topological superconductivity has been found by assuming a finite pairing \cite{poyhonen_amorphous_2018,manna_noncrystalline_2022} its appearance in a self-consistent calculation is yet to be demonstrated. 
Engineering the interactions to obtain a self-consistent topological ground state is a nontrivial problem since Anderson's theorem \cite{Anderson_theory_1959} is strictly applicable only to conventional $s$-wave pairing.

The search for novel topological phases and phenomenology should also incorporate phenomena familiar from crystals. Amorphous topological states have for example not been fully explored in amorphous interacting~\cite{Prodan2019,Kim2022,Cassella2022,Grushin2020}, driven, or non-hermitian systems~\cite{Manna2022}.

Our focus on amorphous systems has necessarily left aside other non-crystalline solids. Disordered crystals, alloys, quasicrystals, fractals, and moiré heterostructures all present exciting opportunities to apply and extend numerous concepts presented here.

In summary, amorphous solids are central to fundamental science and technology. If we aim to establish a full theory of topological matter that is technologically useful it seems unavoidable to consider amorphous materials as the largest subset of non-crystalline solids. We are confident that research in this direction will bring a deeper understanding of condensed matter, as well as novel and interesting phenomenology.

\section{Acknowledgements}
We thank S. Franca and J. Schirmann for discussions and related collaborations. We thank F. Hellman for her invaluable help in understanding amorphous solids. A.G.G. and Q.M. acknowledge financial support from the European Union Horizon 2020 research and innovation program under grant agreement No. 829044 (SCHINES). A.G.G. is also supported by the European Research Council (ERC) Consolidator grant under grant agreement No. 101042707 (TOPOMORPH). P.C. was primarily funded by the US Department of Energy, Office of Science, Office of Basic Energy Sciences, Materials Sciences and Engineering Division under Contract No. DE-AC02-05-CH11231 (NEMM program MSMAG). D.M.S. is supported by an FPU predoctoral contract from Spanish MCIU No. FPU19/03195. J.D.H is supported by the Swedish Research Council (VR) through grant numbers 2019-04736 and 2020-00214.

\bibliographystyle{epl2.bst}
\bibliography{eplpersp.bib}

\end{document}